\documentclass[twocolumn]{aastex6}

\usepackage{hyperref}
\usepackage{amsmath}

\usepackage{color}
\usepackage{soul}

\usepackage{graphicx}
\usepackage{epstopdf}
\makeatletter
\def\env@matrix{\hskip -\arraycolsep 
	\let\@ifnextchar\new@ifnextchar
	\array{*{\c@MaxMatrixCols}c}}
\makeatother
\usepackage{blindtext}

\shorttitle{Pebble Dynamics in Protoplanetary Disks}
\shortauthors{Shadmehri, Khajenabi, and Pessah}

\begin{document}

\title{On the dynamics of pebbles in protoplanetary disks with magnetically-driven winds}

\author{M. Shadmehri$^{1}$, F. Khajenabi$^{1}$, M. E. Pessah$^{2}$}
\affil{
$^1$ Department of Physics, Faculty of Science, Golestan University, Gorgan 49138-15739, Iran\\
$^{2}$ Niels Bohr International Academy, Niels Bohr Institute, Blegdamsvej 17, DK-2100 Copenhagen \O{}, Denmark}
\email{m.shadmehri@gu.ac.ir} 

\begin{abstract}
We present an analytical model to investigate the production of pebbles and their radial transport through a protoplanetary disk (PPD) with magnetically driven winds. While most of the previous analytical studies in this context assume that the radial turbulent coefficient is equal to the vertical dust diffusion coefficient, in the light of the results of recent numerical simulations, we relax this assumption by adopting effective parametrisations of the turbulent coefficients involved in terms of the strength of the magnetic fields driving the wind. Theoretical studies have already pointed out that even in the absence of  winds, these coefficients are not necessarily equal, though its consequences regarding pebble production have not been explored. In this paper, we investigate the evolution of the pebble production line, the radial mass flux of the pebbles and their corresponding surface density as a function of the plasma parameter at the disk midplane. Our analysis explicitly demonstrates that the presence of magnetically-driven winds in a PPD leads to considerable reduction of the rate and duration of the pebble delivery. We show that when the wind is strong, the core growth in mass  due to  the pebble accretion is so slow that it is unlikely that a core could reach a pebble isolation mass during a PPD lifetime. When the mass of a core reaches this critical value, pebble accretion is halted   due to core-driven perturbations in the gas. With decreasing wind strength, however, pebble accretion may, in a shorter time, increase the mass of a core to the  pebble isolation mass.
\end{abstract}


\section{Introduction}

The formation of  planetary systems is a long-standing problem and has attracted considerable interest over recent years \citep[][]{Blum2008,Armitage11,Winn2015, Nayakshin17}. Depending on the physical properties of protoplanetary disks (PPDs), several mechanisms have been proposed to explain planet formation in these complex systems. Gravitational instability, for instance, may eventually lead to  giant planet formation at the  outer parts of a PPD \citep[][]{Boss97,Rafikov2005, Forgan13,Boss17}, whereas the core accretion theory is believed to be the dominant planet formation mechanism at the inner regions of a PPD \citep[][]{Pollack96,Matsuo2007,Boley09}. Both of these planet formation theories, however, have their own weaknesses and are subject to controversy.

In the core-accretion theory, for instance,  core mass requirement to trigger the accretion of gas is debated because the growth time-scale of a massive solid core can be longer than the lifetime of the PPDs \citep{Rafikov2011}. The typical size of solids that can be accreted to form planetary cores is rather controversial \citep{Kobayashi2011,Alibert2017}. Various physical factors, including disk metallicity, the location in the disk, and the accretion rate of solids, can dramatically affect the efficiency of planet formation by the core-accretion scenario \citep{Fortier2013,Piso2014}. Planet formation by gravitational instability, on the other hand,  depends on the cooling efficiency of a disk which is still a challenging issue \citep[][]{Meru11,Rice15,Takahashi16}. Under typical conditions in the gravitationally unstable disks, however, \cite{Kratter} showed that  disk-born fragments are able to grow beyond the deuterium burning planetary mass limit. This implies that wide orbit gravitational instabilities typically produce stars, not planets. During recent years a mixed planet formation scheme known as tidal-downsizing has also been developed which supplements giant gaseous protoplanet formation at the outer parts of a PPD due to gravitational instability and simultaneous migration of this newly formed protoplanet and sedimentation of the grains deep inside it to form a solid core \cite[][]{Helled08,Nayakshin10,Nayakshin15}. 

While the accretion of grains onto a planetary embryo is not very efficient, it has been shown that for  particles with sizes from centimeter to meter, the accretion rate onto a  planetary embryo is enhanced due to the drag force exerted by the ambient gas  \citep{Ormel10,Lambrechts12,Visser16}. Centimeter to meter-sized  particles, which  migrate efficiently  in a PPD, are called {\it pebbles}. Therefore, it is very important to understand how micron-sized particles are able to grow to this range of size and how these pebbles once formed are delivered throughout the PPD \citep[][]{Lamb14,Levison15,Moriarty15, Chambers16, Ida16, IGM16, Sato16,Krijt16,Picogna}. \cite{Lamb14} (hereafter; LJ2014) investigated the production of pebbles and their transport in a gaseous disk with  power law profiles for the surface density and temperature. Their analytical effort showed that the pebble production line spreads outwards and the radial pebble flux through the disk has a weak time dependence. \cite{IGM16} investigated the radial dependence of the pebble accretion rate and showed that the outcome of the pebble accretion scenario strongly depends on the physical conditions in the disk, the drag law and the dimension of the accretion model (i.e., 2D or 3D). \cite{Ida16} presented an analytical model for exploring the effect of the snow line on the inward flow of pebbles and showed that the  accumulation of particles may occur interior to the snow line due to the sublimation of the migrating pebbles. While most  analytical models for the dynamics of pebbles are based on the steady-state gaseous disk model,  \cite{Bitsch15} studied pebble production and delivery in an evolving gaseous disk model. They found that the  pebble accretion scenario is able to resolve many of the challenging issues about ice and gas giant planet formation in evolving PPDs.

To the best of our knowledge, none of the previous works on pebble formation and delivery through a PPD has considered the potential role of the presence of a wind. There are, however,  a number of observational and theoretical studies that suggest that a PPD can lose its mass and/or angular momentum  through winds or outflows, affecting significantly its dynamical structure \citep[][]{Blandford82,Combet2008,Guilet2014, Bjerkeli2016, Bai16,Wang2017}. The reduction of mass in PPDs by winds has been traditionally  attributed to  photoevaporation  \citep[][]{Hollenbach94,Clarke2001,Ruden2004,Gorti2015}, but in recent years, due to extensive studies of the role of non-ideal magnetohydrodynamics (MHD) effects in PPDs,  magneto-driven winds have been at the forefront of the theoretical developments \citep[][]{Salmeron11,Bai13, Bai16}. While it is commonly believed that photoevaporative winds are active and effective at the outer parts of a PPD, especially during the final stages of its evolution, theoretical arguments show that magnetic fields can also play a key role in a wind launching \citep[][]{Ramsey2011,Salmeron11,Guilet2014, Bai16}. Given the existing observational evidence about PPDs, however, it is difficult to determine which of these mechanisms  dominate. Most of the previous (semi)analytical models for a disk with a magnetically driven wind are constructed based on the same simplifying assumptions of the standard accretion disk model with some modifications due to  mass loss and angular momentum transport by winds \citep[][]{Combet2008,Bai16,Suzuki16}. 

Although the dynamics of dust particles in an evolving PPD with winds driven by photoevaporation can be studied according to the current models made for such systems, in this study we plan to study dynamics of pebbles in a PPD with magnetically driven winds. We use a steady-state  model for the structure of a gaseous disk  with a magnetically-driven wind, as provided by \cite{Hasegawa17}  on the basis of the results of numerical simulations \citep{Fromang13,Simon13,Zhu15}. In this model, the net accretion rate is due to  stresses resulting from  MHD turbulence within the disk and  a magnetically-driven wind. Both of these stresses are parameterized as a function of $\beta_0$, the ratio between gas and magnetic pressure at the disk midplane   \citep[][]{Simon13,Zhu15}. 

\cite{Zhu15}  showed that the vertical diffusion coefficient of dust particles  is actually a function of the parameter $\beta_0$. Using three-dimensional global unstratified MHD simulations, they explored  properties of MRI-driven turbulence  in the ideal MHD case and a non-ideal case with ambipolar diffusion. Dust transport in the ideal MHD simulations can be reproduced with  analytical models \citep{Youdin07}, however, ambipolar diffusion qualitatively changes dust diffusion coefficients. The difference is due to the nature of turbulence in each of these cases. Turbulence in the simulations with ambipolar diffusion exhibits different temporal autocorrelation functions and power spectra comparing to the generated turbulence in the ideal MHD simulations. Efficient vertical dust diffusion, for instance, is found in the  ambipolar diffusion dominated simulations \citep{Zhu15}. Considering how the stress tensor components and dust diffusion coefficient  depend on  the parameter $\beta_0$, some authors  have investigated the structure of an accretion disk with magnetically driven winds in the steady-state and even time-dependent cases \citep{Armitage13,Hasegawa17}. Just recently, \cite{Hasegawa17} constructed a gaseous disk model with magnetically-driven winds by considering both radially and  vertically angular momentum transport via MHD turbulence and disk winds, respectively, and found that a high accretion rate and efficient dust settling in the HL Tau disc can all be explained well with this model. Since the mass accretion in the HL Tau disk occurs at a high rate, the estimated dust thickness is larger than what is inferred using a conventional viscous disc model  \citep{Car,Akiyama,Pinte16}. \cite{Hasegawa17} suggested that HL Tau disc accretion is driven by the disc turbulence and wind launching which then leads to a high accretion rate, whereas the vertical dust scale height is maintained by the turbulence and the dust diffusion. Their model is able to explain  the presence of a thin dust layer in HL Tau disk with a high  mass accretion rate.

A key finding of the  MHD simulations of a disk with magnetically driven winds is that the diffusion coefficient driving radial angular momentum transport  is not necessarily the same as the coefficient mediating the vertically diffusion  of dust particles \citep{Zhu15}. Most of previous analytical works about pebble production and its delivery assume that these coefficients are the same \citep{Lamb14,Ida16,IGM16}. Since the growth rate of  micron-sized particles to the size of pebbles strongly depends on the turbulent coefficient driving angular momentum transport and vertical diffusion coefficient of the particles, and given that in a disk with a magnetically driven wind these coefficient are not the same, it is reasonable to expect that the formation and transport of  pebbles are affected in the presence of winds.

In this paper, we extend the work of LJ2014 on the dynamics of pebbles in a PPD by considering the effect of  magnetically driven winds. In the next section, we first present  the steady-state model for an accretion disk with winds. In Section \ref{Peb-acc}, we explore the effect of the wind on pebble formation and their transport through the PPDs. We  obtain the radial mass flux of the pebbles and their surface density as functions of time and the initial plasma parameter $\beta_0$ at the disk midplane and other model parameters. In Section \ref{core-growth},  we investigate the effect of winds on the accretion rate of the pebbles onto a planetary embryo. We discuss the implications of our findings in Section 5.

\section{Protoplanetary disk model}

In most of the previous analytical models for analyzing the delivery of pebbles in  PPDs, the gas component is described using power law functions of the radial distance. For instance, the most widely adopted gaseous model is the minimum mass solar nebula \citep{Hayashi81} or a modified version of the standard disk model which is constructed considering  physical properties of the PPDs \citep[e.g.,][]{IGM16}.  These solutions are applicable to  PPDs without winds. However, it is  reasonable to expect that the disk structure is significantly modified  due to the mass and angular momentum loss by winds \citep[][]{Suzuki16}. Our goal is not to present a detailed model for the gas component in the presence of winds, but rather to explore the implications of adopting a semi-analytical model motivated by the results of recent numerical simulations.  For instance, it has been shown that the steady-sate disk accretion rate in the presence of the magnetically driven winds can be approximated as \citep{Fromang13,Suzuki16},
\begin{equation}\label{eq:Mdot}
\dot{M}=\frac{2\pi \Sigma_{\rm g} c_{\rm s}^2}{\Omega_{\rm K}} \left ( W_{ r\phi} + \frac{2r}{\sqrt{\pi} H_{\rm g}} W_{ z\phi} \right ),
\end{equation} 
where $W_{ r \phi}$ and $W_{ z\phi}$ are the normalized accretion stresses in the radial and the vertical directions,  defined as \citep{Suzuki16,Hasegawa17} 
\begin{equation}
W_{ r\phi}=\frac{\int_{-H}^{H} (\overline{\rho u_{\rm r} \delta u_{\phi}} - \overline{B_{\rm r} B_{\phi}} /4\pi) dz}{\Sigma_{\rm g} c_{\rm s}^2},
\end{equation}
\begin{equation}
W_{ z\phi} = \frac{\left (\overline{\rho u_{\rm z} \delta u_{\phi}} - \overline{B_{\rm z} B_{\phi}}/4\pi \right )_{z=-H}^{z=H}}{2\rho_{\rm 0g} c_{\rm s}^2}.
\end{equation}
Here $u_i$ and $B_i$ are the $i-$th component of the gas velocity and the magnetic field, respectively, $\Sigma_{\rm g}$ is the gas surface density, $c_{\rm s}$ is the sound speed and $\rho_{\rm 0g}$ is the gas density at the disk midplane. Furthermore, the disk height of the wind base is represented by $H$. The overbar denotes  time-averaged quantities. Thus, $W_{ r\phi}$ represents the radial angular momentum transport via MHD turbulence, whereas $W_{ z\phi}$ can be regarded as the  stress exerted by the wind, which quantifies the angular momentum transport by the wind. The thickness of the gas component is $\Sigma_{\rm g}$ and $H_{\rm g}$. The  disk is rotating with a Keplerian profile, i.e. $\Omega_{\rm K}=\sqrt{GM_{\star}/r^3}$, where $r$ is the radial distance and $M_{\star}$ denotes the mass of the central star. 

Using the above relations in the time-dependent case and considering more detailed physical processes, \cite{Suzuki16} did an extensive analysis of the PPD evolution in the presence of magnetically driven winds and found a large variety of  surface density profiles \citep[also see,][]{Bai16}. Although they appreciated the role of the wind in the enhancement of the dust-to-gas ratio, they did not focus on understanding the role of the winds in dust transport nor pebble production in a PPD.   

The MHD shearing-box simulations show that the accretion stresses $W_{ r\phi}$ and $W_{ z\phi}$ can be fitted by the following relations \citep{Simon13},
\begin{equation}\label{eq:wrphi}
\log W_{r\phi} = -2.2 + 0.5 \tan^{-1} \left(\frac{4.4-\log\beta_0}{0.5}\right),
\end{equation}
\begin{equation}\label{eq:wzphi}
\log W_{ z\phi}=1.25-\log\beta_0 ,
\end{equation}
where $\beta_0$ is the plasma parameter defined as  $\beta_0  = 8\pi P_{\rm mid}/B_{\rm z}^2$, where $P_{\rm mid}$ is the gas pressure at the midplane of the disk and $B_{\rm z}$ is the vertical component of the magnetic field.  In addition, according to the results of \cite{Zhu15}, the normalized vertical diffusion coefficient of dust particles, $\alpha_{\rm D}=D_{\rm z}/(c_{\rm s} H_{\rm g})$, can be approximated by the following relation,
\begin{equation}
\log \alpha_{\rm D}= 1.1 - \log \beta_0 ,
\end{equation}
where $D_{\rm z}$ represents the vertical diffusion coefficient. We note that the general trend of $\alpha_{\rm D}$ is more closer to the vertical stress  coefficient $W_{ z\phi}$ instead of following the radial stress coefficient $W_{ r\phi}$. Despite various efforts to constrain the radial evolution of the net vertical magnetic field threading the disk \citep{Lubow94,Rothstein2008,Beckwith2009,Lovelace2009,Okuzumi2014,Bai2017}, there are significant uncertainties about this problem which should be addressed in future works. However, following a previous phenomenological approach \citep[][]{Armitage13}, we assume that the midplane plasma parameter $\beta_0$ is constant.

Using equation (\ref{eq:Mdot}), and for a given  accretion rate, the surface density becomes
\begin{equation}\label{eq:Sigma-Main}
\Sigma_{\rm g}=\frac{\dot{M} \Omega_{\rm K}}{3\pi c_{\rm s}^2} \frac{1}{\alpha_{\rm eff} (\beta_{0}, r)},
\end{equation}
where the effective turbulent coefficient is defined as
\begin{equation}\label{eq:alphaeff}
\alpha_{\rm eff}=\frac{2}{3} \left ( W_{ r\phi} + \frac{2r}{\sqrt{\pi} H_{\rm g}} W_{ z\phi} \right ).
\end{equation} 
With the above definition for the turbulent coefficient, the main equation (\ref{eq:Sigma-Main}) resembles  the classical one dimensional viscous disk model.  The accretion rate in a viscous disk model is written as
\begin{equation}\label{eq:mdot-class}
\dot{M}=\frac{6\pi}{r\Omega} \frac{\partial}{\partial r} (r^2 \Omega \nu \Sigma_{\rm g})
\end{equation}
where $\nu$ is the kinematic viscosity. Various mechanisms can be proposed for generating a disk turbulence, however, the kinematic viscosity is obtained via some further assumptions without discussing the true origin of disc turbulence. It has generally been assumed that the radial-azimuthal component of the stress tensor is proportional to the gas pressure where the constant of proportionality is the dimensionless parameter $\alpha_{\rm SS}$ \citep{SS73}. The kinematic viscosity, thus, can be written as $\nu=\alpha_{\rm SS} c_{\rm s} H_{\rm g}$. Since the dynamics of dust particles is dependent on the disc properties, the turbulent viscosity parameter appears into expressions that describe motion of a dust particle such as its relative velocity to the background gas or other dust particles. Upon comparing equations (\ref{eq:Mdot}) and (\ref{eq:mdot-class}), one can easily show that $W_{r\phi}=(3/2)(\nu \Omega/c_{\rm s}^2)$ when wind does not exist. Therefore, the conventional viscosity parameter $\alpha_{\rm SS}$ can be written as $\alpha_{\rm SS}=(2/3)W_{r\phi}\simeq W_{r\phi}$. Using this result, our effective viscosity parameter becomes $\alpha_{\rm eff}=(2/3) W_{r\phi}\equiv \alpha_{\rm SS}$ when angular  momentum removal by wind is neglected.  The advantage of writing the main equation (\ref{eq:Sigma-Main}) similar to the classical viscous disk model is to include the contribution of wind angular momentum transport using an effective viscosity parameter. The viscosity parameter $\alpha_{\rm SS}$ is generally treated as a fixed input parameter, however, the above formulation shows that once this parameter is given as equation (\ref{eq:alphaeff}), the structure of a disk with winds can still be described within the framework of the standard disk model. It is a simplified approach, however, we think its simplicity enables us to explore the role of angular momentum transport by wind and disk turbulence using a single parameter. Upon introducing the turbulent viscosity as equation (\ref{eq:alphaeff}) and replacing it with the usual parameter $\alpha_{\rm SS}$ into the standard disk equations, one can then investigate disc quantities in the presence of winds. 

For a given temperature profile, we can determine the surface density using the above equation. Following most of the previous studies, we also assume that the temperature follows a power law profile of the radial distance, i.e.,  
\begin{figure}
\includegraphics[scale=0.6]{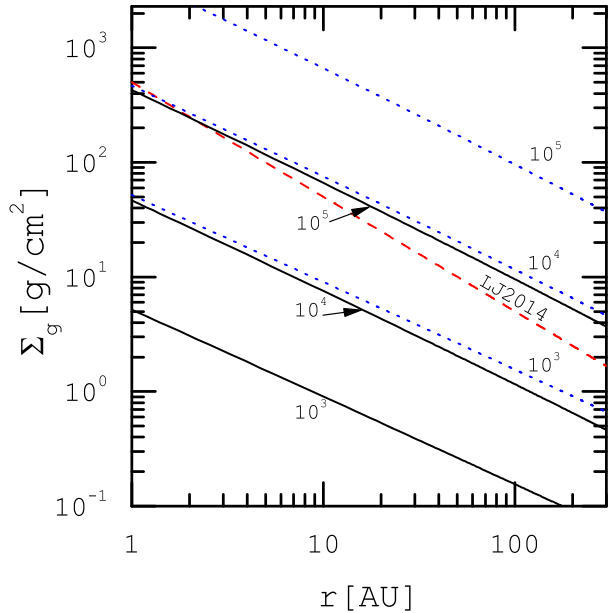}
\caption{Radial profile of the gas surface density for different values of the midplane plasma parameter $\beta_0$ (solid). The central mass is $M_{\star}=M_{\odot}$ and each curve is labelled by the value of $\beta_0$.  Solid lines (black)  correspond to the accretion rate  $\dot{M}=10^{-8}$ M$_{\odot}$/yr, whereas surface density profiles for $\dot{M}=10^{-7}$ M$_{\odot}$/yr are shown by dotted lines (blue).  For  comparison, the surface density profile considered in LJ2014 is shown by a dashed line (red).}\label{fig:f1}
\end{figure}
\begin{equation}
T=T_0 \left ( \frac{r}{r_0} \right )^{-q},
\end{equation}
where $T_0 =100$ K, $r_0 =1$ AU and $q=1/2$ \citep[][]{Alexander2004}. Although this temperature profile is not the most general case and a detailed energy balance for the thermodynamics of the disk has not been considered in the present work, its simplicity  enables us to investigate the role of  magnetically driven winds on  pebble delivery without being involved with the complexities due to the thermodynamics of the disk. Figure \ref{fig:f1} displays our adopted gas surface density profile for different values of the parameter $\beta_0$ and the accretion rate $\dot{M}$. Each curve is marked by the corresponding value of $\beta_0$. The solid lines (black) represent disk surface density for the accretion rate $10^{-8}$ M$_{\odot}$/yr and dotted lines (blue) display surface density profile for $10^{-7}$ M$_{\odot}$/yr. For a given accretion rate,  as the net magnetic field strength increases (i.e., plasma parameter $\beta_0$ becomes smaller), the surface density  reduces significantly. The gas surface density profile of LJ2014, i.e. $\Sigma_{\rm g} = 500 {\rm g} {\rm cm}^{-2}(r/{\rm AU})^{-1}$, is shown by a dashed line in Fig. \ref{fig:f1}.

We note that  the presence of the wind is included only via its role of  transporting angular momentum but excluding the possibility of wind-driven mass-loss. The accretion rate is, thereby, a fixed given input parameter and it does not vary with the distance due to the wind mass-loss. A more realistic model, which is beyond the scope of the present study, incorporates angular momentum removal and mass-loss by the wind \citep{khajenabi18}.  Since the accretion rate in the present disc model  is a given value, as a wind becomes stronger, surface density reduces further and this trend is attributed mainly to the wind angular momentum removal.

It should be pointed out that disk surface density equation (\ref{eq:Sigma-Main}) has already been introduced  by \cite{Hasegawa17}, novel developments only become apparent when this solution is implemented to investigate pebble dynamics in the subsequent sections.

\section{Pebble Dynamics}\label{Peb-acc}
Using presented  gaseous  disk model, we can investigate  dynamics of the dust particles, including pebbles. We first calculate the growth rate of the micron-sized particles to the pebble-sized. Once the growth time-scale becomes comparable to the drift time-scale, the radial migration of the pebbles becomes efficient. Considering the radial motion of the pebbles, we can obtain the radial mass flux of these particles and their surface density as a function of time. 
\subsection{Growth rate of the dust particles}
The dust surface density is assumed  to be a constant fraction of the gas surface density, i.e., 
\begin{equation}
\Sigma_{\rm d}=Z_0 \Sigma_{\rm g},
\end{equation}
where $Z_0$ is the dust-to-gas mass ratio and is assumed  to be equal to the typical ISM value, i.e., $Z_0 =0.01$ (assumed to be a fixed model parameter). We note, however, that both observational evidence and theoretical arguments suggest that $Z_0$  is not necessarily a fixed given value \citep[e.g.,][]{deGre13,Birnstiel14}. 


Using the above gaseous disk model and following the approach of LJ2014, we investigate pebble production and its delivery in the presence of winds. In doing so, the first step is to estimate the growth rate of dust particles. For simplicity, we assume that a dust particle is a sphere with radius $R$. So,  its growth rate is given by \citep{Ormel2007},
\begin{equation}
\frac{dR}{dt}=\frac{1}{4}\frac{\rho_{\rm 0d}}{\rho_{\rm m}} \Delta v_{\rm t}
\end{equation}
where $\rho_{\rm 0d}$ is the volume density of the particles at the disk midplane and $\rho_{\rm m}$ is the material density of a particle. A key quantity in the above equation is the relative turbulent velocity between particles, i.e.,  $\Delta v_{\rm t}$. 

\subsection{Relative velocity between dust particles}

Understanding the processes that determine the relative velocity between dust particles in a turbulent medium is a long-standing problem in the fluid mechanics community \citep[e.g.,][]{Meek73}.  In the astrophysical context, a pioneering model was developed by \cite{Volk80} with subsequent developments and refinements to determine the relative velocity of the particles in a turbulent gaseous medium \citep{Mark91,Cuzz03,Ormel2007}. Using a Kolmogorov power spectrum, the relative velocity of the particles is obtained in terms of the gas Reynolds number and Stokes number that quantifies the dust-gas coupling strength   \citep{Cuzz03,Ormel2007}. The Reynolds number is defined as ${\rm Re}=L V_L / \nu_{m}$, where $L$ is the largest spatial scale, $V_L$ is the associated velocity and $\nu_m$ is the molecular viscosity. Since available relations for the relative dust velocity  depend on the Reynolds number, it is important to have reasonable estimates for ${\rm Re}$. An approach for determining $V_L$ is based on the $\alpha-$formalism, in which the angular momentum is transported   only due to a turbulent viscosity $\nu_T =\alpha c_{\rm s} H_{\rm g}$. Under theses circumstances, we have $V_L =c_{\rm s} \sqrt{\alpha} (\Omega_L / \Omega_{\rm K})^{1/2}$ and $L = H_{\rm g} \sqrt{\alpha} (\Omega_{\rm K} / \Omega_L)^{1/2}$, where $\Omega_L$ is the large-eddy frequency  \citep[see Eqs. (1) and (2) in][]{Cuzzi2001}. 
Defining the Stokes number as ${\rm St}=(\rho_{\rm m} R/\rho_{\rm 0g} c_{\rm s}) \Omega_{\rm K}$ (Epstein regime) and assuming that $\Omega_L \simeq \Omega_{\rm K}$,  the relative velocity in the $\alpha-$disk framework becomes \citep{Ormel2007} 
\begin{equation}\label{eq:del-v}
\Delta v_{\rm t}=(3\alpha {\rm St})^{1/2} c_{\rm s}.
\end{equation}

In obtaining equation (\ref{eq:del-v}) above, we used the $\alpha-$formalism for estimating ${\rm Re}$. There is, however, another approach to determining $V_L$ using the turbulent kinetic energy per unit mass $k$ and an associated $\alpha_k$. \cite{Cuzzi2001} provided a detailed comparative study between the two approaches above for estimating ${\rm Re}$. Following the second approach, they showed that without imposing any restriction on $\Omega_{\rm K}$ and $\Omega_L$, the Reynolds number becomes ${\rm Re}=\alpha_k c_{\rm s} H_{\rm g}/\nu_m$. We, therefore, can still use equation (\ref{eq:del-v}) above, by simply substituting $\alpha=\alpha_{\rm k}$. Invoking energy balance between the energy released due to the accretion and the turbulent kinetic energy dissipated, \cite{Cuzzi2001} showed that
\begin{equation}\label{eq:alpha-k}
\alpha_k = \frac{3\dot{M} \xi_T}{2\pi c_{\rm s} H_{\rm g} \Sigma_{\rm g}},
\end{equation}
where $\xi_T < 1$ is the turbulent energy conversion efficiency. Using equation (\ref{eq:Sigma-Main}) for the disk surface density and equation (\ref{eq:alpha-k}), we obtain $\alpha_{k}=(9/2) \alpha_{\rm eff} \xi_T\simeq \alpha_{\rm eff}$. This suggests that we can substitute $\alpha \equiv \alpha_{\rm eff}$ into equation (\ref{eq:del-v}) to obtain the relative velocity between dust particles in a PPD with magnetic winds.

It is worth noticing that in order to relate $\alpha-$models with the underlying physical processes driving the turbulence it is necessary to rely on  
numerical simulations. In the context of the shearing box, the $\alpha$ parameter can be obtained in the saturated state in turbulent disks subject to the  
magnetorotational instability (MRI) or gravitational instability.
Having said this, however, most previous studies on pebble dynamics, including LJ2014, relied on the $\alpha-$formalism for describing disk turbulence. Magnetically-driven winds, on the other hand, contribute to the disk turbulence and the resulting mass accretion rate. Our disk model encapsulates this key feature too.

We provided some physical motivations for introducing  $\alpha_{\rm eff}$ to incorporate disk wind effects within the framework of the $\alpha-$formalism, however, available non-ideal MHD disk simulations with magnetically-driven winds are useful to address this issue.
\cite{Simon13}, for instance, investigated  a portion of a PPD threaded by a net vertical magnetic field using  a set of shearing box simulations. They found that there is a strong correlation between the strength of the imposed vertical magnetic field and the generated disk turbulence. For sufficiently strong field, however, the main accretion driving mechanism was found to be due to wind launching and not MRI turbulence. They also reported  the associated viscosity coefficient using performed simulations in their Table 1 which are in good agreement with our effective viscosity coefficient $\alpha_{\rm eff}$. \cite{Simon13} found that the viscosity coefficient due to the combined effects of MRI turbulence and wind launching is $\alpha\simeq 0.077, 0.03$ and 0.005 for $\beta_0=10^3, $ $10^4$ and $10^5$, respectively (see their models AD30AU1e3$-$AD30AU1e5). In Figure \ref{fig:f2}, we present the radial profile of $\alpha_{\rm eff}$ for different values of $\beta_0$. Although the spatial dependence of $\alpha_{\rm eff}$ can not be contrasted directly with these shearing box simulations, our effective viscosity parameter at 100 AU is $\alpha_{\rm eff}\simeq 0.1,$ $0.03$ and $0.004$ for $\beta_0=10^3, $ $10^4$ and $10^5$, respectively.

In another approach, \cite{Zhu15} studied ideal and non-ideal MRI-driven turbulence using global and local simulations. They also found that the emergence of winds is a consequence of considering non-ideal terms, including ambipolar diffusion \citep[see also,][]{Gressel}.  They found that disk turbulence is strongly suppressed when ambipolar diffusion is included and it has a longer correlation time for the vertical velocity. They also calculated the power spectrum characterizing the turbulence in both the ideal and the non-ideal MHD cases (see their Figure 9) and the corresponding dust diffusion coefficients (see their Figure 10). The resulting radial and vertical  dust diffusion coefficients in the ideal MHD runs were found in good agreement with the analytical relation of \cite{Youdin07}, who computed these coefficients in a gaseous background with a power spectrum similar to Kolmogorov turbulence.  In the non-ideal runs, however, \cite{Zhu15} found that the dust diffusion coefficients for the particles with ${\rm St}\gtrsim 1$ exhibit noticeable discrepancy from the analytical calculations \cite{Youdin07}. Furthermore, \cite{Zhu15} also computed the viscosity coefficient based on their non-ideal MHD simulations (see their Table 3). Although the value of the viscosity coefficient according to their study is smaller than ours, we suggested that the relative velocity $\Delta v_t$ can be calculated in terms of the effective viscosity coefficient using equation (\ref{eq:del-v}). 

\subsection{Growth time of the dust particles}

\cite{Youdin07} found a  relation between  the thickness of the gaseous disk, $H_{\rm g}$, and the thickness of the dust component, $H_{\rm d}$, which can be written as
\begin{equation}
H_{\rm d}\simeq H_{\rm g} \sqrt{\frac{\alpha_{\rm D}}{\rm St}}.
\end{equation}
This relation is valid so long as the particles are well-coupled to the gas (${\rm St}<1$).

If we define the growth time-scale  of a dust particle as $t_{\rm g}=R/\dot{R}$, then  it  becomes
\begin{equation}\label{eq:tg}
t_{\rm g}=\frac{4}{\sqrt{3} \epsilon_{\rm g} Z \Omega_{\rm K}} \left [ \frac{\alpha_{\rm D}(\beta_0 )}{\alpha_{\rm eff}(\beta_{0}, r)}\right ]^{\frac{1}{2}},
\end{equation}
where the parameter $\epsilon_{\rm g}$ is introduced to control the growth efficiency of the  particles. Using equation (\ref{eq:tg}) for the growth rate of a pebble, one can easily obtain the time $t$  needed for  a particle to grow from its initial size ($R_{\rm initial}$) to its final size ($R_{\rm drift}$). Thus, following LJ2014, we obtain
\begin{equation}\label{Time}
t=\frac{4}{\sqrt{3} \epsilon_{\rm d} Z_0 \Omega_{\rm K}} \left [ \frac{\alpha_{\rm D}(\beta_0 )}{\alpha_{\rm eff}(\beta_{0}, r)} \right ]^{\frac{1}{2}},
\end{equation}
where 
\begin{equation}
\epsilon_{\rm d}=\epsilon_{\rm g} \left [ \ln \left ( \frac{R_{\rm drift}}{R_{\rm initial}} \right ) \right ]^{-1},
\end{equation}
and we have $\ln (R_{\rm drift}/R_{\rm initial}) \simeq 10$. In our model, the dust sticking efficiency is set to $\epsilon_{\rm g} =0.5$.

\subsection{Dominant size of the pebbles}

After time $t$ the growth time-scale becomes comparable to the drift time-scale and a pebble forms which starts its journey towards the central star. The radial drift velocity of a particle is given by \citep{Weiden77,Nakagawa86},
\begin{equation}\label{eq:vr}
v_{\rm r}=-2\frac{{\rm St}}{{\rm St}^2 +1} \eta v_{\rm K},
\end{equation}
where $v_{\rm K}=r\Omega_{\rm K}$ is the Keplerian velocity and $\eta$ is given by
\begin{equation}\label{eq:eta}
\eta = -\frac{1}{2}\left(\frac{H_{\rm g}}{r}\right)^2 \frac{d \ln P}{d \ln r}.
\end{equation}
Therefore, the radial drift time scale is defined as $t_{\rm drift}=r/v_{\rm r}$. If we set the growth time-scale equal to the radial drift time-scale, i.e.,  $t_{\rm g}=t_{\rm drift}$, and in the case with ${\rm St}\lesssim 1$, the dominant size of the pebbles is obtained:
\begin{equation}\label{eq:Stp}
{\rm St}_{\rm p}=\frac{\sqrt{3}}{8}\frac{\epsilon_{\rm p}}{\eta} \frac{\Sigma_{\rm p}}{\Sigma_{\rm g}} \left [ \frac{\alpha_{\rm eff}(\beta_0 , r)}{\alpha_{\rm D}(\beta_0)}\right  ]^{\frac{1}{2}},
\end{equation}
where $\Sigma_{\rm p}$ is the pebble surface density, and, $\epsilon_{\rm p}$ is the coagulation efficiency between pebbles, following LJ2014, we assume to be  $\epsilon_{\rm p}=\epsilon_{\rm g}=0.5$. We note that equation (\ref{eq:Stp}) is obtained using disc quantities with smooth power law profiles. Thus, one can not for example have pressure bumps trapping particles in the disc, then the drift-growth equilibrium breaks down. Particle trapping, however, may occur at certain locations in a disc with magnetic winds when angular momentum removal and mass-loss are considered properly \citep{Suzuki16}.

\subsection{Pebble production front}
Our analysis clearly demonstrates that not only the growth rate but also the dominant size of the pebbles depend on the ratio of the radial and the vertical turbulent coefficients. If we set these coefficients equal to each other, then our equations reduce to the LJ2014 analysis. However, as we pointed out above, numerical simulations of  disks with magnetically driven winds  suggest that these coefficients are {\it not} equal. A few previous studies have also mentioned that the turbulent coefficient which is used for estimating the relative velocity of the particles is  not necessarily equal to the global turbulent coefficient for the viscous evolution of the disk \citep{Carrera17}. This theoretical argument, however, to the best of our knowledge, has not been implemented in the analytical models for pebble production to explore the possible consequences. 

\begin{figure}
\includegraphics[scale=0.6]{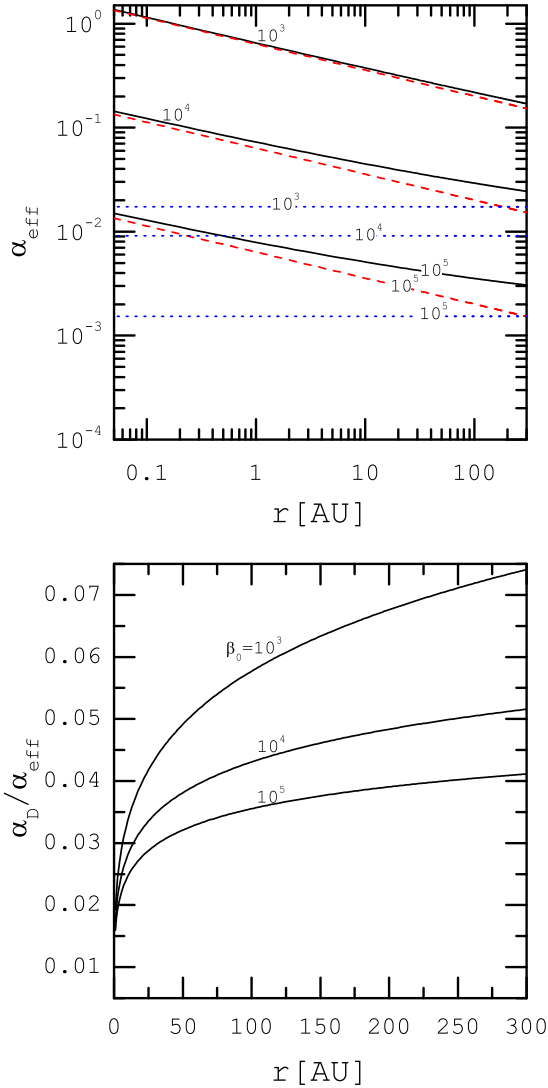}
\caption{Radial profiles of $\alpha_{\rm eff}$ (top) and  the ratio $\alpha_{\rm D}/\alpha_{\rm eff}$ (bottom) for different values of plasma parameter $\beta_0$. In the top plot, solid curves display $\alpha_{\rm eff}$, whereas contributions of the first term (i.e., transport driven by disk turbulence) and the second term (i.e., transport driven by winds) are also shown by dotted (blue) and dashed (red) curves, respectively. Each curve is  labelled by the value of the plasma parameter $\beta_0$. Here, $M_{\star}=M_{\odot}$ and $\dot{M}=10^{-8}$ M$_{\odot}$/yr.}\label{fig:f2}
\end{figure}

Since the ratio $\alpha_{\rm D}/\alpha_{\rm eff}$ plays a key role in our model, we show the radial profiles of $\alpha_{\rm eff}$ and $\alpha_{\rm D}/\alpha_{\rm eff}$ for various values of plasma parameter $\beta_0$ in Figure \ref{fig:f2}. Using equations (\ref{eq:wrphi}) and (\ref{eq:wzphi}), the accretion stresses $W_{r\phi}$ and $W_{z\phi}$ are obtained for a given $\beta_0$, and thereby, the effective turbulent coefficient $\alpha_{\rm eff}$ is obtained from equation (\ref{eq:alphaeff}). The top panel of Fig. \ref{fig:f2} displays not only the effective  viscosity coefficient using equation (\ref{eq:alphaeff}) as solid curves but also contributions of its first term (dotted, blue) and the second term (dashed, red) are shown to have a better insight on the dominant physical mechanism in transporting angular momentum for a given $\beta_0$. As the magnetic field increases (lower $\beta_0$), the contribution of the radial turbulence reduces compared to the angular momentum transport by the winds. For $\beta_0 =10^3$و the dominant angular momentum transport is provided by the winds, whereas for $\beta_0 =10^5$, radial angular momentum transport is most efficient. 

The bottom panel of Fig. \ref{fig:f2} depicts the radial profile of the ratio $\alpha_{\rm D}/\alpha_{\rm eff}$, which is always less than unity for the allowed range of plasma parameter $\beta_0$, however, the ratio increases as $\beta_0$ decreases. This typical trend implies that in a disk with magnetically driven winds, vertical turbulent diffusion is weaker than the radial turbulent coefficient. But as the winds get stronger (lower $\beta_0$), the enhancement of $\alpha_{\rm D}$ is larger than the enhancement of $\alpha_{\rm eff}$, implying a higher value for the ratio $\alpha_{\rm D}/\alpha_{\rm eff}$. Furthermore,  equation (\ref{eq:tg}) implies that the growth time scale is longer than a case with a weak magnetic wind (i.e., smaller $\beta_0$). This general trend eventually affects the location of the pebble production front.

If we set $\alpha_{\rm D}=\alpha_{\rm eff}$, then equation (\ref{eq:tg}) reduces to equation (10) of LJ2014 which gives the pebble production line, $r_{\rm g}$, as a function of time: 
\begin{equation}
r_{\rm g}(t) = 122.7 {\rm AU} \left ( \frac{M_{\star}}{M_{\odot}} \right )^{1/2} \left ( \frac{t}{10^6 {\rm yr}} \right )^{2/3}.
\end{equation}
In the presence of winds, equation (\ref{Time}) can be solved numerically to obtain the time evolution of $r_{\rm g}(t)$. In order to do so, it is convenient to introduce  the following dimensionless quantities:
\begin{equation}
x=\frac{r}{r_0}, M_{1}=\frac{M_{\star}}{M_{\odot}}, \dot{M}_{1}=\frac{\dot{M}}{\dot{M}_{0}} ,t_{6}=\frac{t}{10^{6} {\rm yr}},
\end{equation}
where we take $r_0 = 1 {\rm AU}$ and $\dot{M}_{0}=10^{-8} M_{\odot} {\rm yr}^{-1}$. Thus, we can re-write equation (\ref{eq:alphaeff}) as
\begin{equation}
\alpha_{\rm eff} (\beta_{0}, r)=\frac{2}{3} \left [ W_{ r\phi}(\beta_{0}) + \lambda_{0} M_{1}^{1/2} x^{(q-1)/2} W_{ z\phi}(\beta_{0}) \right ] ,
\end{equation}
where the dimensionless parameter $\lambda_0$ depends on the reference values as $\lambda_0 = (2/\sqrt{\pi}) (r_0 \Omega_0 )/c_{\rm 0s} \simeq 53.6$. Here, we have $\Omega_0 =\sqrt{GM_{\odot}/r_{0}^3}=2\times 10^{-7}$ s$^{-1}$ and $c_{\rm 0s}=\sqrt{k_{\rm B}T_0 / \mu m_{\rm H}}=627.3$ m/s, where $k_{\rm B}$ is  the Boltzmann constant, $\mu=2.1$ is the mean molecular weight and $m_{\rm H}$ is the mass of Hydrogen. We also write the gas surface density equation (\ref{eq:Mdot}) as follows
\begin{equation}\label{eq:sigma-g}
\Sigma_{\rm g}=\Sigma_{0} \frac{ \dot{M}_{1} M_{1}^{1/2}x^{q-\frac{3}{2}}}{W_{ r\phi}(\beta_{0}) +\lambda_{0} M_{1}^{1/2} x^{(q-1)/2} W_{ z\phi} (\beta_{0})},
\end{equation}
where $\Sigma_{0}=\dot{M}_{0} \Omega_0/2\pi c_{\rm 0s}^2 \simeq 5$ g cm$^{-2}$.
With these definitions, the location of the pebble production line, $x_{\rm g}$,  as a function of time is obtained from the dimensionless form of equation (\ref{Time}), i.e.,   
\begin{equation}\label{eq:t6}
t_{6} = \frac{9\times 10^{-4}}{M_{1}^{1/2}} x_{\rm g}^{3/2} \left [ \frac{\alpha_{\rm D}(\beta_0)}{W_{ r\phi} (\beta_0 ) + \lambda_0 M_{1}^{1/2} x_{\rm g}^{(q-1)/2} W_{ z\phi} (\beta_0 )} \right ]^{\frac{1}{2}}.
\end{equation}

\begin{figure}
\includegraphics[scale=0.6]{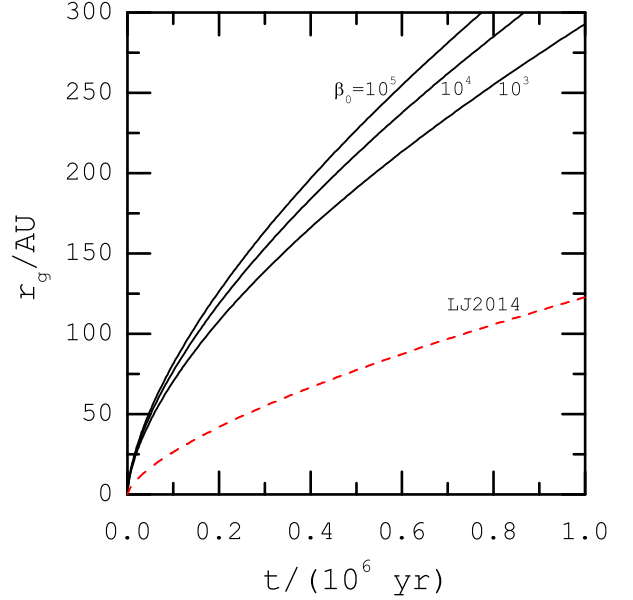}
\caption{The evolution of the pebble production line, $r_{\rm g}(t)$, for different values of the plamsa parameter $\beta_0$ with $M_{\star}=M_{\odot}$, $\dot{M}=10^{-8}$ M$_{\odot}$/yr. Dashed curve (red) shows time evolution of $r_{\rm g}(t)$ based on the LJ2014 model.}\label{fig:f3}
\end{figure}

Figure \ref{fig:f3} displays the time evolution of $r_{\rm g}(t)$ for different values of the parameter $\beta_0$. It shows that the pebble formation front spreads outwards, with an expansion rate that strongly depends on the adopted value for $\beta_0$. As the winds become stronger, the outward propagation  of the  pebble formation line becomes slower. For  comparison, the evolution of $r_{\rm g}(t)$ according to LJ2014 model is shown by a dashed curve (red). We should emphasize that in a weak field case, say, $\beta_0 = 10^{5}$, our time evolution of pebble production front does not converge to the LJ2014 solution. This is simply because the growth rate depends on the ratio $\alpha_{\rm D}/\alpha_{\rm eff}$ which does not tend to unity even in a weak magnetic configuration whereas LJ2014 assumed  this ratio to be always equal to one. Furthermore, our density profile is different from the adopted LJ2014 density profile. However, our treatment enables us to explore the role of magnetic winds on the outward movement of the pebble production line. Slower outward motion of $r_{\rm g}(t)$ in the presence of the winds is understandable in terms of the growth time-scale for pebble formation. We showed that  for a particle with a given initial small size, it takes a longer time for the particle to reach the size at which drift is efficient in the presence of a wind. As the wind becomes stronger, this time scale becomes even longer, and thereby, the pebble formation line spreads at a slower rate. Consequently, Fig. \ref{fig:f3} shows that the pebble formation line reaches  a given radial distance at much later time as the wind gets stronger.

\subsection{Pebble accretion rate}
Now, we can calculate the radial mass flux of the pebbles as
\begin{equation}
\dot{M}_{\cal P} = 2\pi r_{\rm g} \frac{dr_{\rm g}}{dt} \Sigma_{\rm d}(r_{\rm g}),
\end{equation}
where, upon substituting  equation (\ref{eq:sigma-g}) into the above equation, the radial mass flux of the pebbles is given by
\begin{displaymath}
\dot{M}_{\cal P}= (1.2\times 10^{-8} {\rm M}_{\oplus}/{\rm yr}) \frac{dx_{\rm g}}{dt_6} 
\end{displaymath}
\begin{equation}
\times \left [ \frac{\dot{M}_{1} M_{1}^{1/2} x_{\rm g}^{q-(1/2)}}{W_{ r\phi} (\beta_0 ) + \lambda_0 M_{1}^{1/2} x_{\rm g}^{(q-1)/2} W_{ z\phi} (\beta_0 )} \right ].
\end{equation}
Using equation (\ref{eq:t6}), we finally arrive to the following relation for the radial pebble flux, i.e.
\begin{equation}\label{eq:Mdot-P}
\dot{M}_{\cal P}=(8.8\times 10^{-6} {\rm M}_{\oplus}/{\rm yr}) \dot{M}_{1} M_{1}^{3/2} [\alpha_{\rm D}(\beta_0 )]^{-1/2}  x_{\rm g}^{q-1}
\end{equation}
\begin{displaymath}
\times \left \{ \frac{[W_{ r\phi}(\beta_0) + \lambda_0 M_{1}^{1/2} x_{\rm g}^{(q-1)/2} W_{ z\phi}(\beta_0)]^{1/2}}{W_{ r\phi}(\beta_0) + (\frac{7-q}{6}) \lambda_0 M_{1}^{1/2} x_{\rm g}^{(q-1)/2} W_{ z\phi}(\beta_0 )} \right \}.
\end{displaymath}

\begin{figure}
\includegraphics[scale=0.6]{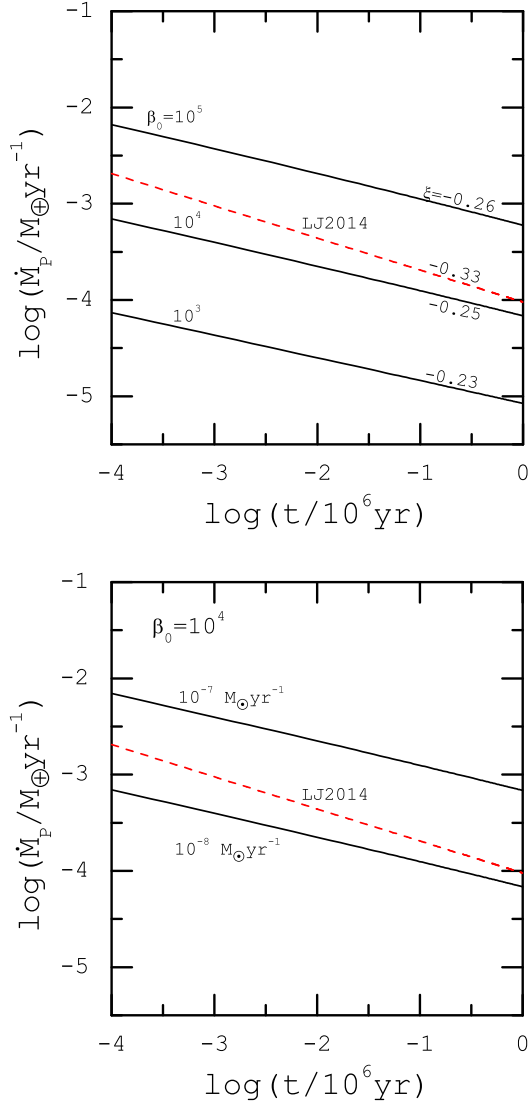}
\caption{Profile of the pebble radial mass flux, $\dot{M}_{\cal P}$, as a function of time for different values of $\beta_0$ and a given accretion rate (top panel) for a fixed plasma parameter and different accretion rates (bottom panel). Here, the central mass is $M_{\star}=M_{\odot}$. The accretion rate in the top panel is $\dot{M}=10^{-8}$ M$_{\odot}$/yr. The numbers beside each curve on the right hand side display the exponent $\xi = d\ln {\dot M}_{\cal P}/d \ln t$. Evolution of the pebble accretion rate based on the LJ2014 relation is shown by a dashed line (red). In the bottom panel, each curve is labeled by the corresponding accretion rate for a fixed plasma parameter $\beta_0 =10^4$. }\label{fig:f4}
\end{figure}

Top panel of Figure \ref{fig:f4} shows the evolution of the radial flux of pebbles for different values of the parameter $\beta_0$ and a fixed accretion rate $\dot{M}=10^{-8}$ M$_{\odot}$/yr. The central mass is $M_{\star}=M_{\odot}$. This figure shows that the time dependence of $\dot{M}_{\cal P}$ can be described as a power law function, i.e. $\dot{M}_{\cal P} \propto t^\xi $, where  the exponent $\xi$ is labeled on the right hand side. To put it more precisely, the stronger the wind, the accretion rate of the pebbles is drastically reduced, however, the exponent $\xi$ has a negligible dependence on the changes of  parameter $\beta_0$. However, the time-dependence of the pebble accretion rate becomes slightly weaker as the winds gets stronger. For comparison, the radial pebble flux based on the LJ2014 model is presented as a dashed line (red). As noted earlier, due to the differences in  the radial and vertical turbulent coefficients, our model with a weak magnetic wind does not reduce to the LJ2014 pebble accretion rate. However, our results show that the presence of the magnetically-driven winds in a PPD can significantly reduce the radial mass flux of the pebbles. 

In  the bottom panel of Figure \ref{fig:f4}, we explore role of the gas accretion rate on the evolution of the radial flux of pebbles. Our adopted disc surface density distribution is directly proportional to the gas accretion rate. A higher accretion rate, therefore, leads to a larger gas surface density at a given radius. Here, the plasma parameter is fixed, i.e. $\beta_0 =10^4$, whereas in addition to our nominal accretion rate, i.e. $10^{-8}$ M$_{\odot}$/yr, a disk with a  higher accretion rate, $10^{-7}$ M$_{\odot}$/yr, is also considered. The pebble radial flux increases with the accretion rate due to enhanced gas surface density. Although the disk surface density is comparable to the LJ2014 profile for an accretion rate $10^{-7}$ M$_{\odot}$/yr, the pebble radial flux is not same as LJ2014. This trend is explainable in terms of the effective viscosity and the vertical dust diffusion coefficients which are not equal in our model.

\begin{figure}
\includegraphics[scale=0.6]{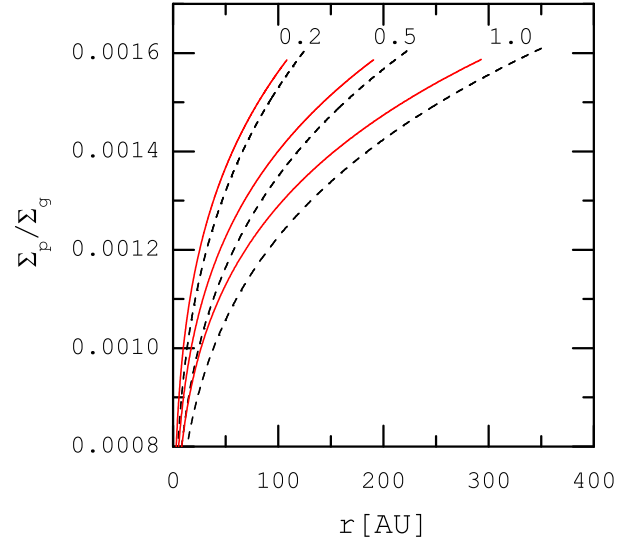}
\caption{Evolution of the ratio $\Sigma_{\rm p}/\Sigma_{\rm g}$ for $\beta_{0}=10^3$ (solid, red) and $10^5$ (dashed, black) and at different times. As in previous figures, $M_{\star}=M_{\odot}$ and $\dot{M}=10^{-8}$ M$_{\odot}$/yr. Each curve is labeld by the corresponding time $t_{6}$.}\label{fig:f5}
\end{figure}

\subsection{Pebble surface density}
We now  calculate the surface density of the pebbles. From equations (\ref{eq:vr}) and (\ref{eq:Stp}), the following radial velocity can be obtained  
\begin{equation}
v_{ r}=\frac{\sqrt{3}}{4} \epsilon_{\rm p} \frac{\Sigma_{\rm p}}{\Sigma_{\rm g}} \left [ \frac{\alpha_{\rm D}(\beta_0 )}{\alpha_{\rm eff}(\beta_{0}, r)} \right ]^{\frac{1}{2}} v_{\rm K},
\end{equation}
where $\Sigma_{\rm p}$ is the surface density of the pebbles and we assumed that ${\rm St}<1$. Therefore, the mass conservation of the pebbles gives the surface density $\Sigma_{\rm p}$ as
\begin{equation}\label{eq:Surf-Peb}
\Sigma_{\rm P}=\left ( \frac{2\dot{M}_{\cal P} \Sigma_{\rm g} }{\sqrt{3} \pi \epsilon_{\rm p} rv_{\rm K} }  \right )^{\frac{1}{2}}\left [\frac{\alpha_{\rm D}(\beta_0 )}{\alpha_{\rm eff}(r,\beta_0 )} \right ]^{\frac{1}{4}},
\end{equation}
where we can substitute the pebble accretion rate from equation (\ref{eq:Mdot-P})  and the gas surface density from equation (\ref{eq:sigma-g}), the above equation. We do not write here the final result of these replacements which are tedious, however, the profile of the pebble surface density evolution is explored in Fig.  \ref{fig:f5}. Two extreme configurations are considered:  a disk with a strong wind (i.e., $\beta_0 = 10^3$, solid curves, black) and a disk with a weak wind (i.e., $\beta_0 = 10^5$, dashed curves, red). Each curve is marked with its elapsed time $t_6$. We note that pebble surface density is shown beyond pebble production line and the curves are limited to the pebble production line. Although we did not consider the reduction in mass through the wind, this figure shows that as the power of the wind increases, the surface density of the pebbles is slightly increased compared to a disk with a weak wind. LJ2014 found the following equation for the ratio of pebble and gas surface densities, i.e.,
\begin{equation}
\left(\frac{\Sigma_{\rm p}}{\Sigma}\right)_{\rm LJ2014}=7.8\times 10^{-4} \left (\frac{t}{10^{6} {\rm yr}} \right )^{-\frac{1}{6}} \left (\frac{r}{\rm AU} \right )^{\frac{1}{4}}.
\end{equation}
This ratio of the surface densities is smaller than what we have found by a factor of 10.

\begin{figure}
\includegraphics[scale=0.6]{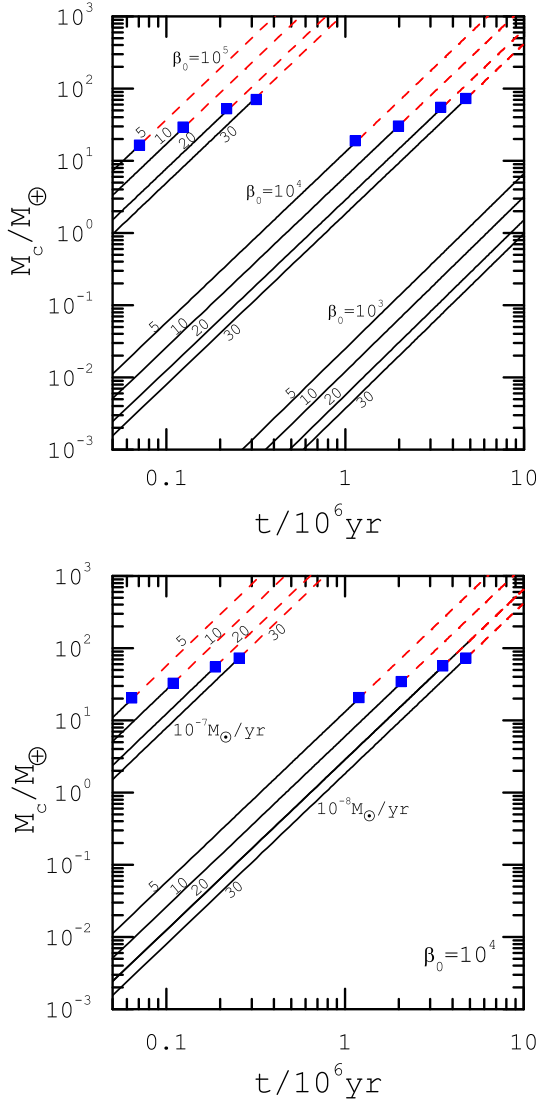}
\caption{Top panel displays mass of a core as a function of time for different values of $\beta_0$ and at different radial distances with a given accretion rate $\dot{M}=10^{-8}$ M$_{\odot}$/yr. The central mass is $M={\rm M}_{\odot}$ and each curve is marked by the corresponding value of the distance in AU. The blue squares indicate when the mass of a core has reached its critical mass and the pebble accretion is stopped afterwards. The red dashed line extrapolates  the growth rate  if the halt of accretion is neglected. Bottom panel is similar to top panel, but for a given plasma parameter, $\beta_0 = 10^4$, and   different accretion rates.}\label{fig:f6}
\end{figure}

\section{Core growth}\label{core-growth}
Obtained the surface density and the radial mass flux of the pebbles, we are now able to calculate the mass accretion rate onto the core of a planetary embryo  with mass $M_{\rm c}$. In the case of ${\rm St}<1$, the accretion rate $\dot{M}_{\rm c}$ can be written as \cite[][]{Lambrechts12},
\begin{equation}
\dot{M}_{\rm c}=2 \left (\frac{\rm St}{0.1} \right )^{\frac{2}{3}} r_{\rm H} v_{\rm H} \Sigma_{\rm p},
\end{equation}
where $r_{\rm H}=r(M_{\rm c}/3M_{\star})^{1/3}$ is the Hill radius, and we have $v_{\rm H}=r_{\rm H} \Omega_{\rm K}$. Using equations (\ref{eq:eta}) and (\ref{eq:Stp}), the above equation becomes
\begin{displaymath}
\dot{M}_{\rm c}=5.79\times 10^{-2} \left (\frac{M_{\star}}{M_{\odot}}\right )^{\frac{1}{12}} \left (\frac{r}{\rm AU} \right )^{\frac{8q-7}{12}} \left (\frac{\Sigma_{\rm g}}{100 {\rm g} {\rm cm}^{-2}} \right )^{\frac{1}{6}}
\end{displaymath}
\begin{equation}\label{eq:Mc-t}
\times \left (\frac{\alpha_{\rm eff}}{\alpha_{\rm D}} \right )^{-\frac{1}{12}} \left (\frac{M_{\rm c}}{M_{\oplus}} \right )^{\frac{2}{3}} \left (\frac{\dot{M}_{\cal P}}{M_{\oplus} {\rm yr}^{-1}} \right )^{\frac{5}{6}} \hspace{0.2cm} \frac{{\rm M}_{\oplus}}{ {\rm yr}}.
\end{equation}

If we substitute for the radial mass flux of the pebbles from equation (\ref{eq:Mdot-P}) and the pebble surface density from equation (\ref{eq:Surf-Peb}), then the growth rate of a core  is obtained as a function of time and the remaining model parameters which can be used to examine the evolution of a core mass due to  pebble accretion.

Equation (\ref{eq:Mc-t}) is, in fact, a first order differential equation in terms of the mass of a core that can be easily integrated to obtain the evolution of the core mass. With some care we see that the time dependence of the right hand side of this  equation appears only through the mass accretion rate $\dot{M}_{\cal P}(t)$. We have already shown that the time dependence of the pebble radial mass flux  can be well approximated  by a power law relation, i.e. $\dot{M}_{\cal P}(t)  \propto t^{\xi}$, where the constant of the proportionality strongly depends on the parameter $\beta_0$, and, the exponent $\xi$ has a relatively weak dependence on the plasma parameter. Using equation (\ref{eq:Mc-t}), thus, we find the following relation, i.e.,
\begin{equation}
M_{\rm c} \simeq  0.1 \left ( \frac{\beta_0}{1000}\right )^{\frac{11}{4}} \left (\frac{r}{\rm AU} \right )^{-1} \left ( \frac{t}{10^6 {\rm yr}}\right )^{3+2.5\xi} {\rm M}_{\oplus},
\end{equation}
where   the exponent $(3+2.5\xi)\simeq 2.3 $  has a relatively weak dependence with the plasma parameter. However, as shown in Fig. \ref{fig:f6}, the growth rate of a core mass is strongly dependent on the parameter $\beta_0$. 

The top panel of Figure \ref{fig:f6} shows the mass of a growing core as a function of time for different values of plasma parameter, a fixed accretion rate, $\dot{M}=10^{-8}$ M$_{\odot}$/yr, and various locations through the disk. The numbers labeling each curve are the radial distance considered in AU. It is clearly seen that the mass of the core increases with time, although the growth rate depends on the wind strength. \cite{Lambrechts2014} showed   that when the mass of a growing core reaches a critical mass $M_{\rm cr}$, then the accretion of the pebbles is halted.  They estimated the critical mass as  a function of the radial distance: $M_{\rm cr} \simeq 20 (r/5{\rm AU})^{3/4}$  ${\rm M}_{\oplus}$ \citep{Lambrechts2014}. More disk parameters, however, have been incorporated in the most recent estimates of the critical mass \citep{Bitsch18,Ataiee18}. While \cite{Bitsch18} generally confirmed  findings of \cite{Lambrechts2014}, results of \cite{Ataiee18} suggest that the pebble critical mass depends on the disk turbulence strength. Our estimate of the critical mass, however, is based on the proposed  relation by \cite{Lambrechts2014}. In Figure \ref{fig:f6}, we identified the time at which the mass of a core  reaches its critical mass by the blue squares, and the continuation of the growth process by ignoring this halt in accretion is extrapolated by the dashed curves (red). When the wind is strong, it can be seen that over the course of 10 million years, core growth due to  pebble accretion occurs so slowly  that it can not reach its critical mass. But with decreasing wind power, the mass of a growing core reaches a critical mass over a shorter time period. 

Although our nominal value for the accretion rate is $\dot{M}=10^{-8}$ M$_{\odot}$/yr, we also displayed gas surface density profile for $\dot{M}=10^{-7}$ M$_{\odot}$/yr in Figure \ref{fig:f1}. Since our gas density profile is directly proportional to the gas accretion rate, the disk surface density increases with the accretion rate for a fixed plasma parameter. The gas surface density dependence of the pebble dynamics has been studied by LJ2014 as illustrated in their Figure 5. They found that pebble accretion rate strongly decreases  with a slight reduction in the disk surface density. We can now explore whether a reduced pebble accretion rate in a strongly magnetized disk is a consequence of having a very low surface density. In Figure \ref{fig:f4}, we explored the role of the accretion rate on the pebble radial flux in a magnetized disk with a gas surface density comparable to that adopted in LJ2014.  

 We have showed that our disk model with $\beta_0 =10^5$ and $\dot{M}=10^{-8}$ M$_{\odot}$/yr is comparable  to LJ2014 density profile, however, pebble growth in this weakly magnetized configuration, as discussed above, is not necessarily  similar to LJ2014. This can be easily understood in terms of the coefficients $\alpha_{\rm eff}$ and $\alpha_{\rm D}$ which are not comparable even in the weakly magnetized discs. This parameter study is generalized to investigate  pebble growth in  a disk with a stronger magnetic field  but comparable gas surface density as in LJ2014.  The surface density of a strongly magnetized disk with $\beta_0 =10^3$, however, becomes comparable to LJ2014, if the accretion rate is adopted around $10^{-6}$ M$_{\odot}$/yr which is not consistent with observed values. We, therefore, consider a case with $\beta_0 =10^4$ and $\dot{M}=10^{-7}$ M$_{\odot}$/yr which leads to a gas surface density comparable to LJ2014. In the bottom panel of Figure \ref{fig:f6},  the plasma parameter is $\beta_0 =10^4$, but different accretion rates are considered. This figure shows that the growth of a core is enhanced when a higher accretion rate is adopted.

When a core reaches its pebble isolation mass, the  gas is able to be accreted onto the core and form an extensive gaseous envelope. Considering our finding that, in the presence of a strong  wind, the mass of a core hardly reaches its critical value, it is unlikely that the formation of giant planets with extensive gas envelopes under these conditions is possible, at least within the pebble accretion scenario. The final fate of such cores will be ice planets that contain little quantities of gas in their envelopes. But in a PPD with a weak wind, a growing core is able to reach its isolation mass and eventually create giant planets. These theoretical speculations depends on other physical processes, such as the  duration of the wind launching and the radial extent of the regions with wind in a PPD. More sophisticated models, however, are needed to address these complicated issues.

\section{Discussion and Conclusions}
We investigated pebble growth and its transport in a PPD with magnetically-driven winds using a simplified steady state model for the gas component. Using results of previous numerical simulations about the behavior of the radial accretion stress and the vertical dust diffusion  \citep[][]{Simon13,Zhu15}, we found that the growth rate of micron-size particles to the pebble-sized strongly depends on the ratio $\alpha_{\rm D}/\alpha_{\rm eff}$, which, in our model, is a function of the midplane plasma parameter $\beta_0$ and the radial distance. 

In our implemented magnetized gaseous disk model, even in a case with weak magnetic winds, contrary to most of the previous (semi)analytical works in this context, this ratio does not tend to unity. The turbulence in a PPD is highly anisotropic and it is unlikely  that turbulence will have the same role in the dynamics of particles in the vertical and radial directions. Transport of the dust particles in a turbulent medium, like a PPD, is an interesting topic and has attracted considerable attention over recent years \citep[e.g.,][]{Turner2010}. Although the dust sub-layer can be subject to different kinds of instabilities,   such as streaming instability \citep[][]{Youdin2005,Johansen2007}, to maintain its thickness, the vertical transport of dust particles is commonly described using the diffusion approximation \citep[][]{Voelk1980,Cuzzi93,Dullemond2004}. We also checked that dust layer in a disk with a high plasma parameter is not thinner than the streaming instability supported minimum of $H_{\rm d}/H_{\rm g}\sim 0.01$ \citep{Yang14,Yang17}.

Under the steady-state approximation and some simplifying assumptions \citep[][]{Youdin07}, the ratio of the thickness of dusty disk and gaseous disk becomes equal to $(\alpha_{\rm D}/{\rm St})^{1/2}$, where ${\rm St}<1$. \cite{Youdin07}  pointed out that the diffusion coefficient $\alpha_{\rm D}$ is {\it not} necessarily equal to the  turbulent coefficient $\alpha$. Numerous numerical simulations have been done over recent years to  investigate dust dynamics  in  PPDs. For instance, \cite{Fromang2009} investigated dust settling in a disk with ideal MHD driven turbulence and found that there is a disagreement between the results of their simulations and simple theoretical expectations,  unless the diffusion coefficient varies vertically in proportion to the square of the local turbulent vertical velocity fluctuations. In another effort, \cite{Bai2010} investigated the dynamics of  dust particles in the midplane of a non-magnetized PPD with turbulence  driven by the streaming instability. They also found that particle dynamics does not follow simple models that are obtained assuming the equality of the radial and vertical turbulent coefficients. 

It is worth noting that our implemented relation for the relative velocity between dust particles $\Delta v_t$, i.e. equation (\ref{eq:del-v}), is an essential part of the present model to calculate pebble growth rate and its evolution. This relation has been used by previous pebble production models in PPDs without winds, however, our physical arguments and recent developments in the non-ideal MHD disk simulations provided motivations to use this equation in PPDs with winds. In this case, we suggested that the viscosity coefficient in equation (\ref{eq:del-v}) can be replaced by an effective viscosity coefficient, i.e. equation (\ref{eq:alphaeff}) which includes a wind contribution as well. We, however, are aware of the limitations of our treatment in estimating the relative velocity $\Delta v_t$. In our argument to support this replacement, we introduced the parameter $\xi_T$ to denote the efficiency with which the released energy is dissipated by turbulence. Although this parameter should be less than unity, its exact value remains uncertain and it depends on the properties of the turbulence. Furthermore, available relations for the relative velocity between dust particles are presented as integrals where the turbulent energy spectrum is a model input function \citep[e.g., see Eq. (12) in][]{Cuzz03}. The Kolmogorov energy spectrum, thereby, has generally been adopted to obtain equation (\ref{eq:del-v}) \citep{Ormel2007}. Non-ideal MHD induced turbulence, on the other hand, exhibits a  power spectrum with noticeable differences in comparison to the ideal MHD simulations, requiring 
 a more realistic approach to compute the relative velocity $\Delta v_t$. 
In principle, the obtained power spectrum based on non-ideal MHD disk simulations \cite[see Fig. 9 in][]{Zhu15} could be implemented in the integral relations of \cite{Cuzz03} or \cite{Ormel2007} to calculate the relative velocity $\Delta v_t$. This approach could offer a potential avenue for improving the present model in the future.

Although we did not consider mass loss due to winds, we found that the pebble surface is slightly enhanced compared to a case with weak winds. In fact, the role of the wind appeared only through changes in the radial and the vertical turbulent coefficients, however, during the period of growth and transport of the pebbles, a non-negligible fraction of the mass of the disk can be lost due to the presence of the winds. More sophisticated models are needed for modeling gaseous disk with magnetic driven winds. For instance, \cite{Bai16} presented a global disk model with magnetic winds in which both angular momentum and mass loss by winds have been considered consistently. \cite{Suzuki16} investigated the  evolution of a PPD with magnetically driven winds within the framework of the  standard accretion disc model \citep{SS73}. The loss of mass and  angular momentum by disk winds are incorporated using parameterized relations constrained by MHD simulations and global energy budget of a disc. \cite{Suzuki16} showed that radial profiles of the disc quantities display a large variety, including an inner surface density with a positive radial slope. These theoretical findings that may affect  the radial drift of dust particles  deserves further studies.  In addition, dust particles, especially smaller ones, may be uplifted to the surface layers of a PPD, and be removed from the disk through the wind. Although the details of these mechanisms are not clear \citep[][]{Miyake2016}, it seems necessary to consider this key physical process  in the next steps to improve the present study.

Based on the present study, in which enhancement of the dust-to-gas mass ratio due to the wind  was not considered, the presence of the magnetically-driven wind could reduce propagation rate of pebble production line   and this delay the onset of the core growth. Although it is not clear which parts of a disk are susceptible to  the wind launching, if the wind is assumed to be active over the entire  disk, then the efficiency of a planetary embryo growth by  pebble accretion can be greatly reduced in the presence of winds. It is unlikely that there will be wind throughout the entire disk. Just recently, \cite{Nolan17} investigated steady-state disc-wind solutions by considering all non-ideal effects and found that disc winds occur over a limited radial extend. In addition, the location of the wind-launching region and its size were found strongly dependent on the mass accretion rate, magnetic field strength and the disc surface density profile.  However, theoretical studies \citep[e.g.,][]{Bai13} and observational evidences \citep[e.g,][]{Bjerkeli2016} have shown that especially in the inner regions of a PPD such magnetically-driven winds can be very active and even operate as the dominant angular momentum transport mechanism. Since the pebble production line spreads outwards, at least in the inner region of a PPD where is the most likely place for wind launching,  there would be a delay in production and transport of pebbles according to the results of our model. Although we implemented a simple model for pebble formation and its delivery, our results at least can show  the expected trends of pebble dynamics under wind launching conditions in a PPD. 

As the wind becomes stronger, pebble production line propagates with a slower rate and the pebble radial mass flux reduces. Rate of the core growth, therefore, reduces in the presence of a strong wind. The plasma parameter plays a vital role in our model, because the wind strength is quantified in terms of this parameter. We found a set of giant planets for $\beta_0 =10^4$ and $\dot{M}=10^{-8}$ M$_{\odot}$/yr, however, formed cores at a larger orbital distance are expected to be more massive. Observational features of HL Tau, on the other hand, suggest that the preferred value is $\beta_0 =10^4$ as discussed in \cite{Hasegawa17}. Our analysis suggests that for the allowed range of the model parameters, pebble growth mostly leads to super-Earths. Current observational studies show that super-Earths are more common in comparison to gas giants which are found rarely (at most around 10 percent of stars). This trend implies that gas giant formation occurs at specific conditions which deserve further study.

The pebble production line in our magnetized disk model propagates to a large  radial  extend. For the allowed range of plasma parameter, therefore, a PPD size is required to be larger than 100 AU which is not consistent with ALMA observations of the PPDs.  In the old Ophiuchus star forming region, for instance, \cite{Andrews2009} did an extensive survey of PPDs and their characteristic size is found between 20 AU to 200 AU.  ALMA observations of 36 PPDs in the Lupus star forming complex exhibit a similar range of the disc size \citep{Tazzari2017}.  Continuous dust replenishment in the PPDs that may alter the mass flux rate as an input of our model or loss of detection efficiency in observations of PPDs are possibilities that may help us to reconcile current observational features of PPDs with our disk wind model. It deserves further works to address these issues.

\begin{acknowledgements}
We are grateful to the referee, Michiel Lambrechts, for a very constructive and thoughtful report that greatly helped us to improve the paper. 
MS and FK thank Golestan University for supporting their research, and, Niels Bohr International Academy and The Niels Bohr Institute for its hospitality and support during part of this work.
The research leading to these results has received funding from the European Research Council under the European Union's Seventh Framework Programme (FP/2007-2013) under ERC grant agreement 306614 (MEP).
\end{acknowledgements}

\bibliographystyle{apj}
\bibliography{reference} 
\clearpage


\end{document}